# Fluid dynamics simulations show that facial masks can suppress the spread of COVID-19 in indoor environments

Ali Khosronejad[1*], Christian Santoni[1], Kevin Flora[1], Zexia Zhang[1], Seokkoo Kang[2], Seyedmehdi Payabvash[3], and Fotis Sotiropoulos[1]

[1] Civil Engineering Department, Stony Brook University, Stony Brook, NY 11704

[2] Department of Civil and Environmental Engineering, Hanyang University, Seoul 04763, South Korea

[3] Department of Radiology and Biomedical Imaging, Yale University, New Haven, CT 06519

* Ali Khosronejad, Civil Engineering Department, Stony Brook University, Stony Brook, NY 11794, Tel: (631) 632-9222, Email: ali.khosronejad@stonybrook.edu

## Abstract

The Coronavirus disease outbreak of 2019 has been causing significant loss of life and unprecedented economical loss throughout the world. Social distancing and face masks are widely recommended around the globe in order to protect others and prevent the spread of the virus through breathing, coughing, and sneezing. To expand the scientific underpinnings of such recommendations, we carry out high-fidelity computational fluid dynamics simulations of unprecedented resolution and realism to elucidate the underlying physics of saliva particulate transport during human cough with and without facial masks. Our simulations: (a) are carried out under both a stagnant ambient flow (indoor) and a mild unidirectional breeze (outdoor); (b) incorporate the effect of human anatomy on the flow; (c) account for both medical and non-medical grade masks; and (d) consider a wide spectrum of particulate sizes, ranging from 10 µm to 300 µm. We show that during indoor coughing some saliva particulates could travel up to 0.48 m, 0.73 m, and 2.62 m for the cases with medical-grade, non-medical grade, and without facial masks, respectively. Thus, in indoor environments either medical or non-medical grade facial masks can successfully limit the spreading of saliva particulates to others. Under outdoor conditions with a unidirectional mild breeze, however, leakage flow through the mask can cause saliva particulates to be entrained into the energetic shear layers around the body and transported very fast at large distances by the turbulent flow, thus, limiting the effectiveness of facial masks.

**Keywords:** Fluid dynamics of coughing; Saliva particulate transport; Face masks; SARS-CoV-2 virus; Turbulence; Large eddy simulation.

## 1. Introduction

As the Coronavirus Disease of 2019 (COVID-19) pandemic, caused by severe acute respiratory syndrome coronavirus [2] (SARS-CoV-2), is spreading around the globe, it is causing devasting loss of life, threatens to topple national health care systems, and wreaks havoc on the global economy. The World Health Organization (WHO), the United States Centers for Disease Control (CDC), and other public health organizations are recommending social distancing and the use of facial coverings as two of the most important weapons to curb the spread of the pandemic [1-5]. However, recommended social distancing guidelines vary considerably with the WHO recommending at least 1 m while the CDC recommending at least 1.8 m of separation [5]. There is now mounting evidence that such guidelines are based on outdated scientific studies carried out decades ago [6], which, among others, have overlooked the physics of respiratory emissions, the range of droplet size distribution such emissions produce, the turbulent flow processes that transport respiratory droplets and the potential effects of ambient wind conditions on such transport [6-14]. Several experimental and computational studies are now beginning to emerge, seeking to fill this major knowledge gap and yield improved science-based physical distancing guidelines (see Refs. 5 and 7) for excellent recent reviews).

Experimental studies have suggested, for instance, that under certain conditions, pathogen-bearing droplets emitted during violent expiratory events (coughing and sneezing) can travel up to 7-8 m [8-11]. Dbouk and Drikakis [12,13] employed a statistically stationary Reynolds-averaged Navier-Stokes (RANS) computational fluid dynamics (CFD) model to simulate droplet transport during coughing and showed that while saliva droplets from coughing travel less than 2 m under stagnant ambient air, they could travel very fast and for several meters downwind when a mild ambient breeze is present as a result of turbulence transport. In another recent study, Cummins et al. [14] investigated the dynamics of various size individual droplet dispersion in a source-sink pair flow filed using a simplified numerical model and showed that unlike the large droplets, the intermediate-sized droplets have a minimal dispersion range. Studies are also emerging, seeking to quantify the effect of facial masks as a means to limit the transport of pathogen-bearing droplets. Fischer et al. [9] designed a low-cost experimental method to probe the respiratory droplet transmission of regular speech through different facial masks. The results showed that N95 respirators, surgical, and cotton masks strongly reduce respiratory droplet count, whereas the bandana did not show the same effectiveness. Verma et al. [10] emulated coughs using the vaporized liquid mixture to visualize the effectiveness of different mask materials and designs. Their study demonstrated that uncovered coughs could spread farther than the recommended 1.8 m social distance, while homemade and commercial facial masks can significantly curtail the speed and spreading range of the saliva particulates, even though with some leakage around mask edges.

Coughing and sneezing give rise to jet-like flows whose Reynolds number is approximately 10,000 [11]. At this Reynolds number, the flow becomes fully three-dimensional and turbulent and is dominated by energetic coherent structures spanning a range of scales. While the critical role of turbulence in transporting expiratory droplets is well recognized [11-13], studies resolving the scales of turbulence generated during an expiratory event as it interacts with facial coverings and/or ambient turbulence have yet to report in the literature. Moreover, no studies have been reported that take into account the presence of the body of the human that generates the droplets. While for expiration in stagnant air such simplification may be adequate, when the ambient wind is present, neglecting the presence of the human is a gross over-simplification as the body generates additional turbulence and sheds energetic coherent structures that could significantly impact the transport of respiratory droplets by the flow. In this work, we employ a high-fidelity large-eddy simulation (LES) CFD model to study the transport of droplets of various sizes generated during coughing with and without facial masks and under varying ambient flow conditions. Our work is significant because we: 1) employ unprecedented numerical resolution to carry out the LES, using over three orders of magnitude finer grids that previous studies (e.g., see Refs. 12 and 13) and resolving the energy of turbulent eddies down to 0.5 mm in size; 2) incorporate into the simulation the full three-dimensional geometry of facial coverings; 3) take into account the entire anatomy of the human that generates the cough; and 4) report simulations both for stagnant ambient air and a mild breeze.

We employ the open-source Virtual Flow Simulator massively parallel CFD code, which has been developed by our group and applied in the past extensive to a wide range of biological, environmental, and engineering turbulent flows [15-23]. Our method employs second-order finite differencing numerics, the dynamic Smagorinsky model for modeling the unresolved subgrid scales of turbulence in the LES, the sharp interface curvilinear immersed boundary (CURVIB) method [15] to simulate the three-dimensional anatomy of the human body, and a new CURVIB-based albeit diffused interface immersed boundary approach to incorporate the presence of facial masks and their different porosities. The saliva particulates are treated as active tracers in stratified turbulent flow, and the stratification effect is accounted for using the Boussinesq approximation in the LES equations of fluid motion [15-18]. Simulations on computational grids with over 650 million nodes are performed under both stagnant air (referred to as indoor) and a unidirectional mild breeze that blows forward at a mean-flow velocity of U = 4.5 m/s (referred to as outdoor) conditions. Two different facial masks, including a medical grade N95 and a non-medical grade facial mask, are incorporated in the simulations, which also incorporate the 3D anatomy of a generic human body. Overall, five scenarios are simulated and analyzed: (a) indoor cough without a facial mask; (b) indoor cough with a medical-grade facial mask; (c) indoor cough with a non-medical grade facial mask; (d) outdoor cough without a facial mask; and (e) outdoor cough with a non-medical grade facial mask.

## 2. Numerical Model

### 2.1. Governing equations

We employ a fully coupled aerodynamics and saliva transport model capable of carrying out LES of stratified, turbulent flows with the presence of arbitrarily complex geometries, such as a face, mouth, or facial masks [15]. We solve the spatially-filtered continuity and Navier-Stokes equations with the Boussinesq assumption to simulate the incompressible, stratified, turbulent flow of dilute air-saliva mixture using a second-order accurate finite-difference fractional step approach, which satisfies the discrete continuity equation to machine zero at each time step. The Navier-Stokes equations with the Boussinesq assumption in non-orthogonal, generalized, curvilinear coordinates $\{\xi^i\}$, and compact tensor notation (repeated indices imply summation) are as follows ($i = 1,2,$ or 3 and $j = 1, 2,$ and 3):

$$J\frac{\partial U^j}{\partial \xi^j} = 0 \tag{1}$$

$$\frac{1}{J}\frac{\partial U^j}{\partial t} = \frac{\xi_l^i}{J}\left[-\frac{\partial(U^j u_l)}{\partial \xi_j} + \frac{1}{\rho_0 Re}\frac{\partial}{\partial \xi^j}\left(\mu\frac{\xi_l^j \xi_l^k}{J}\frac{\partial u_l}{\partial \xi^k}\right) - \frac{1}{\rho_0}\frac{\partial}{\partial \xi^j}\left(\frac{\xi_l^j p}{J}\right) - \frac{1}{\rho_0}\frac{\partial \tau_{lj}}{\partial \xi^j} + \frac{\bar{\rho}-\rho_0}{\rho_0}g\left(\frac{\delta_{i3}}{J}\right) + f_l\right] \tag{2}$$

where $\xi_l^i$ are the transformation metrics, $J$ is the Jacobian of the transformation, $U^i$ is the contravariant volume flux, $u_i$ is the Cartesian velocity component, $p$ is the pressure, $\tau_{lj}$ is the Reynolds stress tensor for the LES model, $\delta_{ij}$ is the Kronecker delta, $\mu$ is the dynamic viscosity, $g$ is the gravitational acceleration, $\rho_0$ is the background density (the density of air in our case), $\bar{\rho}$ is the density of the air-saliva mixture, and $f_l$ ($l = 1, 2,$ and 3) is the body force introduced by the facial mask computed using the mask model, as described below. Our method employs second-order finite differencing numerics, the dynamic Smagorinsky model [25] for modeling the unresolved subgrid scales of turbulence in the LES. The pressure Poisson equation is solved using an algebraic multigrid acceleration along with generalized minimal residual method (GMRES) solver and the discretized momentum equation is solved using a matrix-free Newton-Krylov method. The density of the air-saliva mixture is modeled as

$$\bar{\rho} = \rho_0(1 - \psi) + \rho_s \psi \tag{3}$$

where $\psi$ is the volume fraction of the saliva particulates and $\rho_s$ is the density of saliva particulates, which is considered to be equal to that of the water (=1000 $kg/m^3$). For dilute air-saliva mixture in which the volumetric saliva concentration barely exceeds $O(0.01)$, the saliva particulates concentration is modeled as an active scalar whose transport is computed using the following convection-diffusion equation[17]:

$$\frac{1}{J}\frac{\partial(\rho_0 \psi)}{\partial t} + \frac{\partial}{\partial \xi^j}\langle\rho_0\psi(U^j - W^j\delta_{i3})\rangle - \frac{\partial}{\partial \xi^j}\langle(\mu\sigma_L + \mu_t\sigma_T)\frac{\xi_l^j \xi_l^k}{J}\frac{\partial \psi}{\partial \xi^k}\rangle \tag{4}$$

where $W^j = (\frac{\xi_3^j}{J})w_s$ is the contravariant volume flux of saliva particulates in the vertical direction due to the settling velocity ($w_s$) of the particulates, $\sigma_L$ is the laminar Schmidt number (=1) (21), $\sigma_T$ is the turbulent Schmidt number (=0.75)[21], and $\mu_t$ is the eddy viscosity obtained from the LES model. Finally, it should be noted that in this study, we do not consider the evaporation of saliva during and after coughing, as described in Bourouiba et al.[6]. Although evaporation can play a role in the transport of saliva particulates[6], dropping its effects should not affect the generality of our findings as they pertain to the relative effects of indoor and outdoor transport with and without facial masks. In fact, we could reasonably argue that our simulations may provide more conservative estimates of transport as droplets do not evaporate.

**2.2. Numerical model of the human anatomy and facial masks**

We employ the sharp interface curvilinear immersed boundary (CURVIB) method with wall model reconstruction [15] to simulate the anatomy of the human body (Fig. 1a) while a diffused interface immersed boundary method is used to incorporate the presence of facial masks and their different porosities. Namely, the facial mask is modeled by applying a drag force on the unstructured grid nodes used to discretize the three-dimensional geometry of the mask (see Fig. 1d). The so prescribed force is then distributed to the cartesian grid nodes of the background grid via a smoothed discrete delta function, which is used as the kernel for transferring information between the two grids. That is, the facial mask momentum drag force term (i.e., $f_l$ in Eqn. 2) is given by:

$$f_l = \frac{1}{2}\bar{\rho}\, C_D a(u_i u_i)^{1/2} u_l \Delta(x_j - X_j) \quad (5)$$

where $C_D$ is the drag coefficient, $a$ is the projected area of the facial mask, $u_i$ is the local cartesian velocity vector, and $\Delta$ is the smoothed discrete delta function. The approach we employ for implementing the drag force over the mask geometry is essentially identical to that presented in Yang and Sotiropoulos [26] for prescribing forces over wind turbine structures. The geometry of the two facial masks, with an average thickness of 2.2 mm, were created using an open-source software, Blender (www.blender.org) (Fig. 1b,c). The details of the numerical method and validation studies for jet-like flows, which resemble the cough, are documented extensively elsewhere (e.g., see Refs. 8-16).

## 2.3. Computational details

A computational domain with the dimension of 4.0 m in length, 1.0 m in width, and 2.5 m in height is simulated using the spatial and temporal resolution of 0.5 mm and 0.5 ms, respectively, resulting in over 650 million computational grid nodes. To the best of our knowledge, this is the highest resolution ever reported for the simulation of coughing flow around a human body. The periodic boundary condition was adopted in the spanwise direction. The no-slip boundary condition was used for the bottom boundary representing the ground, while the top and outlet of the domain were prescribed using Neumann outflow boundary conditions. The location and the shape of the mouth opening (i.e., 2.2 cm wide and 1.5 cm high [11-14]. We used the drag coefficients of 600 and 350 in the facial mask model to represent the medical grade and non-medical facial masks, respectively [13]. The numerical model is parallelized using the message passing interface (MPI) communication standard and runs on a parallel high-performance supercomputer. On average, each of the five test cases was simulated using 1000 CPUs of our supercomputing cluster for about 4 months of CPU time to reach the moment when the forward momentum of saliva particulates, and, consequently, their spreading, approached machine zero.

## 3. Results

We report a series of high-fidelity numerical simulation results of the aforementioned five scenarios to examine the effect of facial masks on the transport of saliva during coughs. Multiple saliva particulate sizes of 10 µm, 20 µm, 40 µm, 80 µm, 160 µm, and 300 µm with settling velocities of 0.003 m/s, 0.012 m/s, 0.049 m/s, 0.193 m/s, 0.770, and 2.707 m/s, respectively, were considered in the simulations. The cough pulse considered in the simulations takes 0.24 s to complete [11-13]. It starts from the velocity of zero at $t = 0$ and increases until it reaches a maximum velocity of 8.5 m/s at $t = 0.12$ s and eventually reduces down to zero again at $t = 0.24$ s.

### 3.1. Cough saliva particulates transport under indoor conditions without a facial mask

In Figs. 2 and 3, we illustrate the instantaneous simulation results of a cough without a facial mask and in a room with stagnant air conditions for saliva particulate sizes of 10 µm (Fig. 2), 40 µm (Fig. 3), and 80 µm (Fig. 3). For the sake of brevity, the results obtained for the rest of the particulate sizes are not presented here but can be made available upon request. As seen in Fig. 3e-h, the 80 µm saliva particulates settle down soon after they are emitted, reaching a maximum distance of 1.13 m from the person. The finer particulates, however, travel farther with the 10 µm particulates traveling the farthest away from the person (Fig. 2). The 10 µm saliva particulates constitute less than 5% of the saliva emitted during coughing [11-13], however, they are large enough to carry SARS-CoV-2 viruses. Thus, to be conservative, in the rest of the study, we focus our attention on the simulation results of the 10 µm saliva particulates. We continued the numerical simulations until the forward momentum of the 10 µm saliva particulates reached machine zero (i.e., $10^8$) (Fig. 2). Beyond this time (in most cases at $t > 300$ s), the 10 µm saliva particulates only settle to the ground with a fall velocity of 0.003 m/s. As seen in Fig. 2, without wearing a facial mask the 10 µm saliva particulates ejected during the cough can travel up to 2.62 m in the direction of the cough in a room with stagnant air conditions. Even well after the forward momentum of the cough reaches machine zero, the finer saliva particulates in our simulation (10 µm) are found to stay suspended in the stagnant air for longer than 5 mins as a result of their small settling velocity. According to a recent study by Zhang et al. [24], airborne transmission is the main route for the spread of COVID-19. Thus, our finding regarding the transport of cough saliva particulates under indoor conditions without a mask can have important implications for other people in the same room since the area contaminated with the saliva particulates can extend up to 2.62 m in front and about 0.94 m across the coughing individual for 5 min or longer. The simulated scenarios presented herein provide the baseline to examine the efficacy of facial masks in limiting the spread of saliva particulates, as discussed in the next sections.

**3.2. Cough saliva particulates transport under indoor conditions with facial masks**

In Fig. 4, we plot the instantaneous simulation results of the spreading of 10 µm saliva particulates after coughing with the medical and non-medical grade facial masks. As seen in this figure, both mask types act as effective sinks of forward momentum and limit the spreading of the saliva particulates by both re-directing portions of the cough in the vertical direction and dissipating its forward momentum. Vortices "B" and "T" in Fig. 4m,n mark the vertical re-direction of the cough towards the mask openings at the bottom (anterior neck) (see vortices "B" in Fig. 4m,n) and top (over the top of the nose), respectively. The portion of the cough flow that passes directly through the fabric of the facial masks transporting some of the particulates forward is marked by vortices "P1," "P2," and "P3" in Fig. 4m,n. Since the medical-grade facial mask (Fig. 4a-f,m) has a greater drag coefficient than the non-medical grade mask (Fig. 4g-l,n), the cough flow rate passing through the medical-grade mask seems to be about 24% smaller than that for the non-medical grade facial mask. The effectiveness of the facial masks in dissipating the forward propagation of the free jet-like cough flow can be quantified by comparing the temporal variation of the instantaneous total kinetic energy of the flow for the cases with and without the mask in Fig. 5. As seen in this figure, the total kinetic energy of the cough flow with the medical and non-medical facial masks alike is about one order of magnitude smaller than for the same cough without a facial mask. Thus, our results provide evidence that even non-medical facial masks can be very effective in dissipating the energy of the jet-like cough flow, which helps the saliva particulates to begin to settle before they reach far from the coughing person. After about 485 s and 432 s since the start of the cough, the indoor coughs with the medical and non-medical grade facial masks, respectively, lose their forward momentum and, therefore, only move in the vertical direction settling down. At this time, the computed maximum distance that saliva particulates have traveled away from the person is 0.48 m and 0.73 m for the medical and non-medical grade facial masks, respectively. Compared with the same saliva particulate (10 µm) transport without a facial mask in Fig. 2, the simulation results clearly mark the effectiveness of both types of facial masks (medical and non-medical grades) in limiting the spread of the 10 µm saliva particulates under indoor stagnant air conditions.

### 3.3. Saliva transport during coughing under outdoor conditions

In this section, we simulate the same cough processes without a facial mask and with the non-medical grade facial mask by replacing the background flow condition of the indoor stagnant air with a turbulent background flow condition, which corresponds to a unidirectional mild breeze with mean-flow velocity of 4.5 m/s. These simulations also incorporate and resolve the full anatomical details of the person. Before initiating the cough in the simulation, we first simulate the mild breeze that flows from behind the person and into the computational domain to obtain the fully developed turbulent flow past the human body. As the incoming flow is diverted around and over the body, a complex turbulent flow arises dominated by energetic three-dimensional shear layers shed from the body and a three-dimensional separation zone in the wake of the person (i.e., the front side of the body) (see Fig. 6). Once the turbulent flow around the person reaches statistical equilibrium, we initiate in the simulation the cough process to simulate saliva transport with and without facial masks in an ambient environment with a mild turbulent breeze flow. A snapshots of the so computed instantaneous flow fields are shown in Fig. 6.

As seen in Fig. 7a-e, for the case without a facial mask, the saliva particulates are trapped in the three-dimensional recirculation flow region in front of the person. Since the flow is highly turbulent and fully three-dimensional, however, saliva particulates can randomly escape the recirculation zone and travel forward. Once saliva particulates escape the recirculation zone, their fate is dominated by the background turbulent air flow and, thus, can travel forward rapidly and reach very far distances away from the person. In fact, as seen in Fig. 7e, the saliva particulates can get transported nearly 2.0 m in less than a second.

Furthermore, as seen in Fig. 7f-j, the saliva transport process for the case with the facial mask is almost identical to that of the case without a facial mask. This is because, and as we discussed in the previous section, facial masks exhibit leakage and divert some of the saliva particulates upward toward the top of the head where particulates can escape the recirculation zone as they get entrained by the fast traveling outer turbulent flow. Once this happens and saliva particulates have escaped into the ambient flow, they can travel forward very rapidly, also reaching approximately 2.2 m ahead of the person in less than a second. As seen in Fig. 7j, the transport of the saliva particulates in the case with the facial mask occurs very rapidly and as fast as that of the case without the facial mask. Therefore, our results show that, unlike indoor conditions, the use of facial masks in outdoor conditions (with a unidirectional and forward blowing mild breeze) cannot limit the transport of the fine saliva particulates during the cough as a result of the entrainment of mask leakage flow by the fast traveling ambient turbulent flow.

## 4. Conclusion

We carried out high-fidelity LES on a computational grid that is three orders of magnitude finer than previously reported simulations to study saliva transport during coughing. We conducted five different simulations to examine the efficacy of medical and non-medical grade facial masks in suppressing the spread of saliva particulates during the coughing process under indoor (stagnant background air) and outdoor (unidirectional mild breeze) conditions. For each of the five simulations, we used on average 1000 CPUs for about 4 months of CPU time.

As summarized in Fig. 8, our simulation results for the saliva particulate transport during the cough under indoor conditions showed that, if no facial mask is used, the finer saliva particulates can travel up to about 2.62 m away from the person, while particulates of larger sizes settle down between the person and that distance, in less than 2 min from the start of the cough. Moreover, our simulations also show that under indoor conditions, finer saliva particulates can remain suspended above the ground for several minutes and at a height where they can be inhaled by others or entrained by a ventilation system even after their forward momentum has decayed to zero. Therefore, our results clearly show that under indoor conditions and without a facial covering neither the WHO nor the CDC social distancing guidelines provide adequate protection and need to be revised upward. The use of medical and non-medical facial masks, however, can effectively limit the traveling distance of the finer particulates to about 0.48 m and 0.73 m, respectively, away from the person and under the indoor stagnant air conditions. Therefore, our results show that combining either medical or non-medical grade facial masks with either the WHO or the CDC social distancing guidelines should be very effective in protecting against infection from the spread of virus-carrying expiratory cough droplets in indoor environments.

For outdoor flow conditions and under a mild unidirectional breeze, however, a dramatically different picture emerges. Our simulation results showed that regardless of whether a facial mask is present the fine saliva particulates during the cough can rapidly travel very long distances away from the person in the direction of the breeze in a very short period of time. This is because even when a facial mask is worn the upward vertical movement of the saliva particulates induced by the mask can cause particulates to be entrained by the fast traveling turbulent ambient flow of the breeze near the top of the head of the person causing them to travel forward very fast to reach many meters away from the person in just seconds. Therefore, our results suggest that, while in indoor environments wearing a mask is very effective to protect others, in outdoor conditions with ambient wind flow present wearing a mask might be essential to protect ourselves from pathogen-carrying saliva particulates escaping from another mask wearing individual in the vicinity.

Finally, we should note that because of the high computational cost of the simulations, we limited our scenarios for the outdoor conditions to a unidirectional mild breeze. Our numerical method, however, can be used to computationally investigate a variety of ambient flow conditions as well as other mask types and designs.

**Acknowledgments**

This work was supported by grants from the National Science Foundation (EAR-0120914) and a sub-award from the National Institute of Health (2R44ES025070-02). The first author would like to thank Aram Khosronejad for the proofreading and her contributions to enhance the readability of the paper.

**Data Availability Statement**

The data that support the findings of this study are available from the corresponding author upon reasonable request.

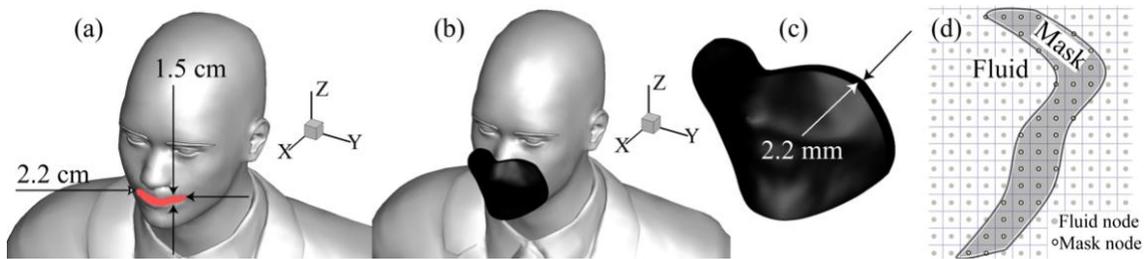

**Figure 1:** Simulated human anatomy and the facial mask from 3D view. (a) shows the face and dimensions of mouth opening (in red) during the cough. (b) shows the human anatomy and the facial mask covering the face and mouth. (c) shows the facial mask. (d) schematic of the fluid and mask nodes of the background grid.

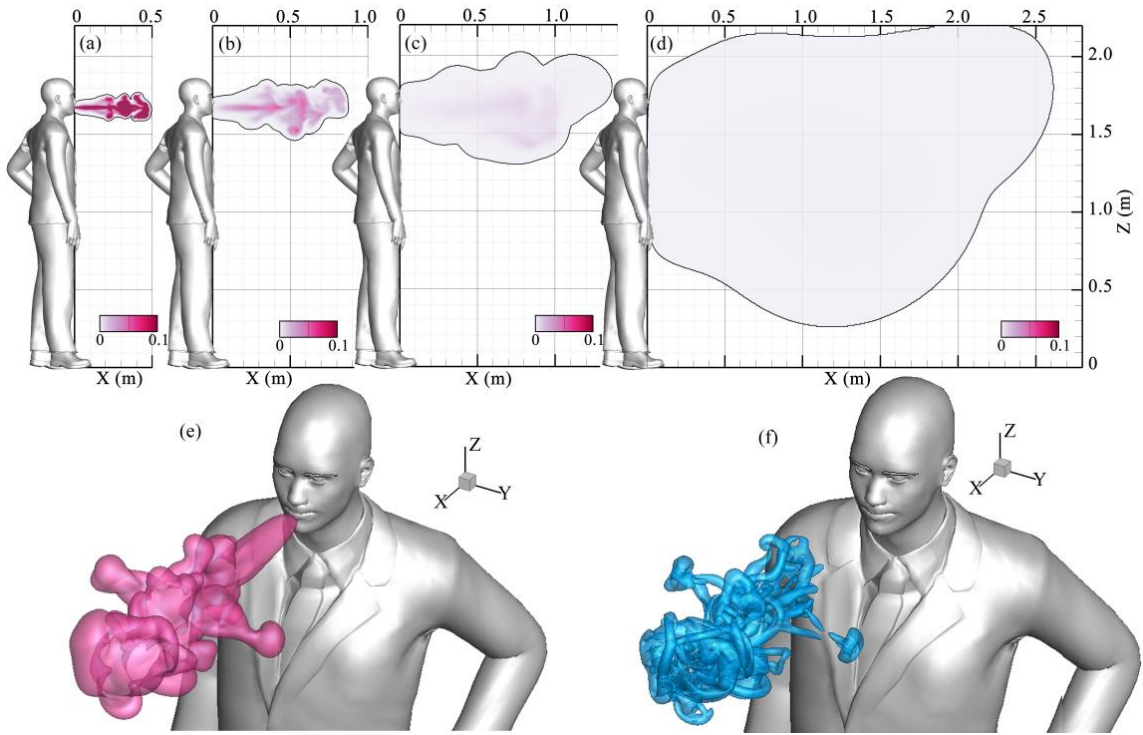

**Figure 2.** Simulated evolution of the 10 µm saliva particulate concentration (volume fraction) after the cough under indoor conditions without a facial mask. (a), (b), (c), and (d) show instantaneous snapshots of the simulated saliva particulate concentration fields after 0.24 s, 1.0 s, 6.0 s, and 310 s, respectively, on the sagittal (or longitudinal) plane. In (e) and (f) we visualize the three-dimensional structure of the cough after 0.5 s in terms of (e) an iso-surface of saliva particulate concentration (=0.01); and (f) an iso-surface of the Q-criterion (19) illustrating the complex vortical structures arising in the turbulent cough jet.

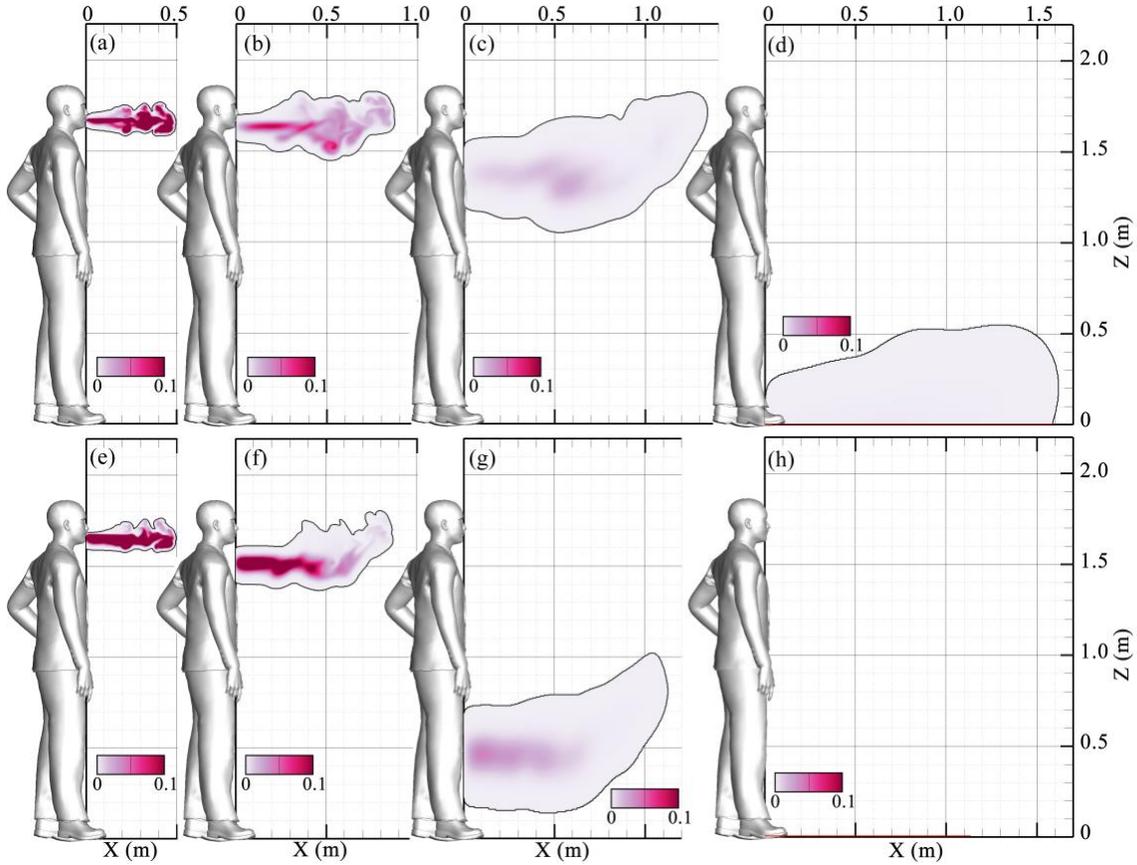

**Figure 3.** Simulated evolution of the 40 µm (top) and 80 µm (bottom) saliva particulate concentration (volume fraction) contours after the cough under indoor conditions without a facial mask. (a,e); (b,f); (c,g); and (d,h) show the simulated saliva particulate concentration fields after 0.24 s; 1.0 s; 6.0 s; and 310 s, respectively, on the sagittal plane.

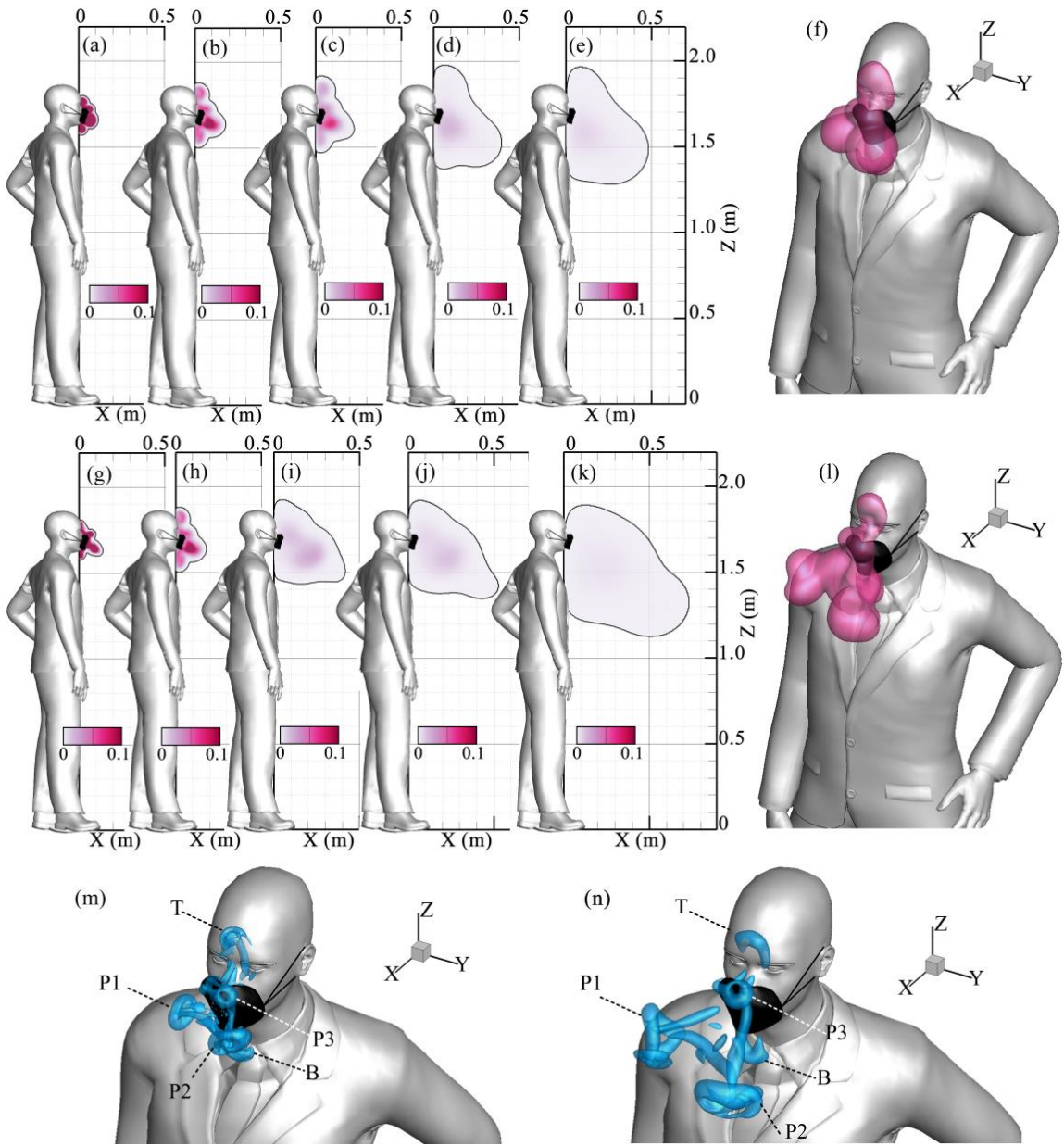

**Figure 4.** Simulated evolution of the 10 µm saliva particulate concentration (volume fraction) after the cough under indoor conditions with the medical (top row) and non-medical (middle row) grade facial masks. (a,g); (b,h); (c,i); (d,j); (e); and (k) show the simulated saliva particulate concentration fields after 0.24 s; 1.0 s; 6.0 s; 200 s; 485 s; and 432s, respectively, on the sagittal plane. (f), (l), (m), and (n) show the cough in 3D after 0.5 s. (f) and (l) visualize the cough using the iso-surfaces of saliva particulate concentration (=0.01) with the medical and non-medical facial masks, respectively. (m) and (n) visualize the cough vortical flow structures using the iso-surfaces of Q-criterion

(19) with the medical and non-medical facial masks, respectively. The "P1" to "P3" vortices in (m) and (n) mark the portion of the cough flow that passes through the fabric of the facial masks, while the "T" and "B" vortices illustrate the cough flow that is re-directed toward the top and bottom of the facial masks, respectively.

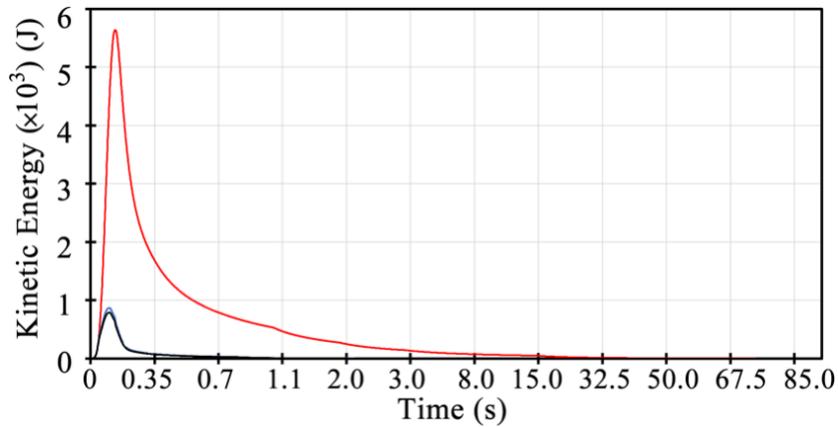

**Figure 5.** Simulated evolution of the total kinetic energy (J) of the cough under indoor conditions with and without the facial mask. Red, black, and blue lines represent the total kinetic energy of cough without a mask, with a medical grade mask, and with a non-medical grade mask, respectively. As seen, medical and non-medical masks reduced the total kinetic energy of the cough and, consequently, its convective force by one order of magnitude. After about a minute the total kinetic energy of all indoor cases approaches machine zero and, thus, we only show 85 s of the kinetic energy variations.

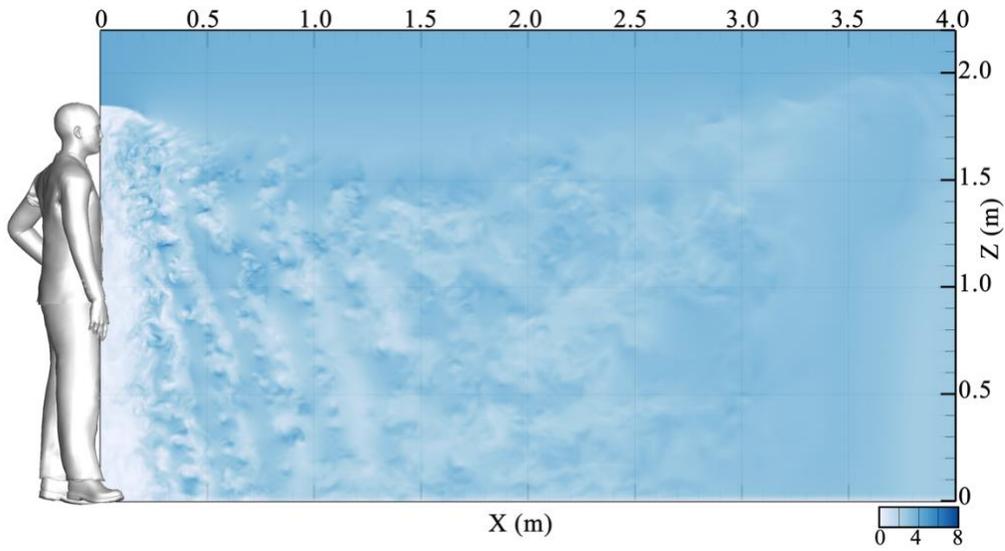

**Figure 6:** Instantaneous contours of velocity magnitude (m/s) of the fully developed flow of the mild breeze around the person on the sagittal plane. The 3D flow field, a 2D snapshot of which is depicted in here, was used as the initial flow condition before starting the cough for both the cases with and without facial masks.

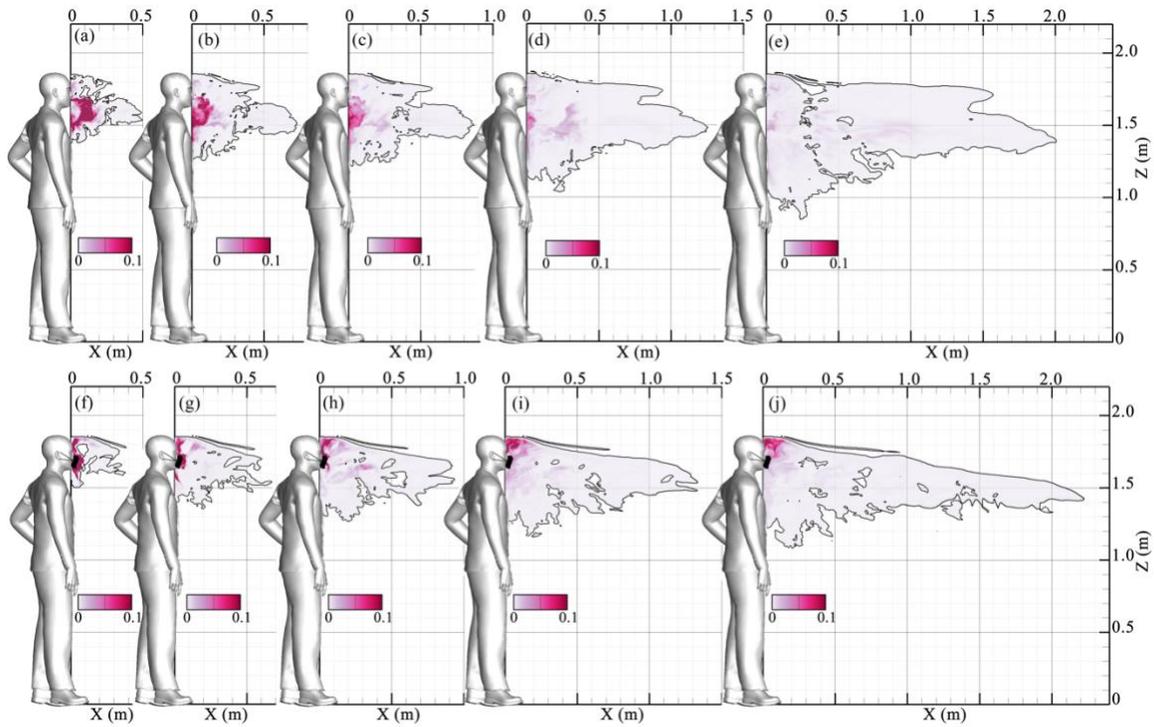

**Figure 7:** Simulated evolution of the 10 µm saliva particulate concentration (volume fraction) after the cough under outdoor conditions (mild breeze) without (top) and with (bottom) the facial mask. (a,f); (b,g); (c,h); (d,i); and (e,j) show the simulated saliva particulate concentration fields after 0.24 s; 0.3 s; 0.4 s; and 0.5 s; and 0.6 s, respectively, on the sagittal plane. The outdoor simulations were stopped after 0.6 s, when the saliva particulates travel approximately 2.0 m and 2.2 m without (top) and with (bottom) the facial mask, respectively.

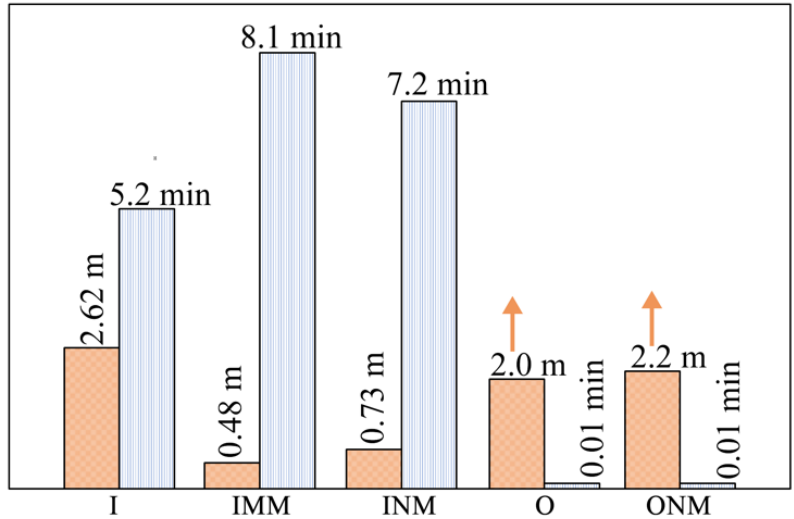

**Figure 8:** Simulated spreading length of the 10 µm saliva particulates (orange bars) and the time (blue bars) it takes to reach that maximum distance. (I) shows the case under indoor conditions without a facial mask. (IMM) and (INM) represent the cases under indoor conditions with medical grade and non-medical grade facial masks, respectively. The outdoor cases without a facial mask and with the non-medical grade facial mask are shown by (O) and (ONM), respectively. As seen, under outdoor conditions, facial masks could limit neither the maximum length of spreading nor the speed of spreading, as saliva particulates travel rapidly 2.2 m away from the person, due to the turbulent flow of the ambient air. (O) and (ONM) simulations were stopped after 0.6 s, when the saliva particulates had traveled approximately 2.0 m and 2.2 m, respectively, away from the person. The orange arrows on the spreading length of (O) and (ONM) cases show that the saliva spreading process continues at $t > 0.6$ s. Given the very fast propagation speed of particulates in these cases, it is clear that they will continue propagating for many meters away from the source before they begin to settle to the ground.